\documentclass[
bibnotes,
amsmath,amssymb,
prl,						 %Modify
twocolumn,                   %Modify
letterpaper,
twoside,
frontmatterverbose,
floatfix,
notitlepage,
]{revtex4-1}

\usepackage{amsmath}
\usepackage{mathtools}
\usepackage{graphicx}% Include figure files
\usepackage{dcolumn}% Align table columns on decimal point
\usepackage{bm}% bold math
\usepackage{multirow}
\usepackage{array}
\usepackage{booktabs}
\usepackage{upgreek}
\usepackage{epsfig}
\usepackage{mathrsfs}
\usepackage{amssymb}
\usepackage{amsbsy}
\usepackage{color}
\usepackage{cancel}
\usepackage{pifont}
\usepackage{marginnote}
\usepackage{float}
\usepackage{tikz}
\usepackage{bbm}
\usepackage{ctable}

\usetikzlibrary{arrows}
\tikzstyle{block}=[draw opacity=0.7,line width=1.4cm]
\usetikzlibrary{positioning}
\usepackage[caption=false]{subfig}
\usepackage{setspace}

\newcommand{\bcen}{\begin{center}}
\newcommand{\ecen}{\end{center}}
\newcommand{\btab}{\begin{tabular}}
\newcommand{\etab}{\end{tabular}}
\newcommand{\bdes}{\begin{description}}
\newcommand{\edes}{\end{description}}

\newcommand{\beq}{\begin{equation}}
\newcommand{\eeq}{\end{equation}}
\newcommand{\bea}{\begin{eqnarray}}
\newcommand{\eea}{\end{eqnarray}}

\newcommand{\half}{\frac{1}{2}}
\newcommand{\bary}{\begin{array}}
\newcommand{\eary}{\end{array}}
\newcommand{\benum}{\begin{enumerate}}
\newcommand{\eenum}{\end{enumerate}}
\newcommand{\bitem}{\begin{itemize}}
\newcommand{\eitem}{\end{itemize}}

\newcommand{\dou}{\partial}

\newcommand{\D}[1]{\mbox{d}{#1}}

\newcommand{\paratitle}[1]{\noindent {\bf #1:}}

\newcommand{\eqn}[1] {eqn.~(\ref{#1})}

\newcommand{\fig}[1]{fig.~\ref{#1}}
\newcommand{\Fig}[1]{Fig.~\ref{#1}}

\makeatletter

\newcommand{\Rmnum}[1]{\expandafter\@slowromancap\romannumeral #1@}
\makeatother

\newcommand{\mylabel}[1]{\label{#1}}

\newcommand{\authorprl}[2]{\author{#1}\email{#2}}

\newcommand{\titlename}{Strange Half Metals and Mott Insulators in SYK Models}
\DeclareMathOperator{\sgn}{sgn}

\newcommand{\qs}{{q_s}}
\newcommand{\qc}{{q_c}}
\newcommand{\qPsi}{{q_\Psi}}
\newcommand{\Lo}{{N}}
\newcommand{\Lc}{{N_c}}
\newcommand{\LPsi}{{N_\Psi}}

\newcommand{\Jc}{{J_c}}
\newcommand{\JPsi}{{J_\Psi}}

\newcommand{\Sigc}{{\Sigma_c}}
\newcommand{\SigPsi}{{\Sigma_\Psi}}
\newcommand{\fx}{f}

\newcommand{\cH}{\mathscr{H}}

\newcommand{\ci}{\mathbbm{i}}

\newcommand{\dens}{{n}}
\newcommand{\NSM}{Strange Half Metal}
\newcommand{\NSMab}{SHM}
\newcommand{\nsm}{strange half metal}

\begin{document}

%\preprint{}
%\preprint{}

\title{\titlename}
\authorprl{Arijit Haldar}{arijit@physics.iisc.ernet.in}
\authorprl{Vijay B. Shenoy}{shenoy@physics.iisc.ernet.in}
\affiliation{Centre for Condensed Matter Theory, Department of Physics, Indian Institute of Science, Bangalore 560 012, India}


\date{\today}


%%%%%%%%%%%%%%%%%%%%%%%%%%%%%%%%%%%%%%%%%%%
\begin{abstract} 
%We study a model consisting of two fermion flavors each of which forms a Sachdev-Ye-Kitaev system with arbitrary $q$-body interactions.
We study a dual flavor fermion model where each of the flavors form a Sachdev-Ye-Kitaev (SYK) system with arbitrary and possibly distinct $q$-body interactions.
 The crucial new element is an arbitrary all-to-all $r$-body interaction between the two flavors. At high temperatures the model shows a strange metal phase where both  flavors are gapless, similar to the usual single flavor SYK model.
  Upon reducing temperature, the coupled system undergoes phase transitions to  previously unseen phases - first, a \nsm\ phase where one flavor remains a strange metal  while the other is gapped, and, second, a Mott insulating phase where both flavors are gapped. At a fixed low temperature we obtain transitions between these phases by tuning the relative fraction of sites for each flavor. We discuss the physics of these phases and the nature of transitions between them. This work provides an example of an instability of the strange metal with potential to provide new routes to  study strongly correlated systems through the rich physics contained in SYK like models.
\end{abstract}

\maketitle

\paratitle{Introduction} Some of the most fascinating open problems of condensed matter physics pertain to gapless many fermion systems -- usually dubbed as ``strange metals''\cite{Varma2002singular} -- which lack single fermion excitations  that are particle like. It is well recognized that understanding the physics of such systems can throw light on a number of important materials such as the cuprate high temperature superconductors\cite{LeeWenNagaosa2006}, and, from a more fundamental perspective, provide a framework for the description of properties of fermionic systems near quantum critical points. The central difficulty in dealing with these issues is the innate non-perturbative character of such systems, compounded by the fact that the traditional phenomenological approaches such as Boltzmann transport theory are not even qualitatively adequate to describe their rich physics.
 
Significant recent progress seen in this area has resulted from  ``holographic'' approaches \cite{HartnollLecHol,Gubser1998gauge,Witten1998anti,Sachdev2010PRL,Zaanen2015Book} which exploit a connection between a quantum field theory living on the boundary of certain spacetimes to gravitational theories in the bulk of these spacetimes. While these approaches have tasted success\cite{SubirGraphene}, interpretation of gravitational results in terms of the boundary condensed matter quantities has not been straightforward. In this regard, a notable recent development is the revival of a solvable model -- which currently goes under the name of Sachdev-Ye-Kitaev (SYK) model -- for strange metals, originally proposed by Sachdev and Ye \cite{SachdevYe1993}, which was shown by Kitaev\cite{Kitaev2015} to have a connection to two dimensional gravity. The SYK model consists of a ``zero dimensional'' system with $N$ sites in which fermions have an ``all to all'' random interaction whose variance provides the basic energy scale. Originally, Sachdev and Ye\cite{SachdevYe1993} and Kitaev\cite{Kitaev2015}, considered two body interactions (Kitaev version of the model also uses Majorana fermions),  since then the model has been generalized to arbitrary $q$-body interactions\cite{Maldacena2016,Polchinski2016}, and a formulation of systems without disorder\cite{Witten2016SYK} has also been made. The solvability of the model is a result of the simplification that arises in the large $N$ limit which singles out a certain set of quantum process which dominate the physics. 

A remarkable aspect of the SYK model is that there is a residual ground state entropy that arises from the structure of low energy many body spectrum and is determined solely by $q$. Several other features, such as the divergent infrared spectral density of states of the gapless fermionic excitations, are also dependent solely on $q$. Sachdev\cite{Sachdev2015PRX} has demonstrated the connection between the SYK model and the thermodynamics of certain black holes. This was done by relating the nonzero ground state entropy of the SYK model with the Bekenstein-Hawking entropy of the black hole on the gravitational side. Further, the SYK model can also teach us about equilibration and thermalization (see Kitaev\cite{Kitaev2015}); the SYK model is known to be ``chaotic'' in that it provides an example of a system with the ``most efficient'' thermal relaxation.  This and much other rich physics of this model has been revealed in recent papers \cite{Maldacena2016,Polchinski2016}, which has added to important results pointed out in the pioneering works\cite{SachdevYe1993,Parcolett1997,Parcolett1998,Parcolett2000,Parcolett2001}.

Apropos the study of strange metals and other condensed matter issues, a new direction has been to employ higher dimensional generalizations of the SYK model to study transport in systems without quasiparticles \cite{Gu2016,Davison2016thermo}. From the perspective of equilibration and relaxation, Banerjee and Altman have recently proposed an SYK like model with two flavors of fermions which provides a system that can be tuned from a chaotic relaxor to a slow scrambler(see \cite{BanerjeeAltman2016}), the former being a strange metal like phase and the latter a Fermi liquid like phase. This model also provides an interesting framework to study transitions between a strange metal to a Fermi liquid phase.

These recent developments motivate intriguing questions, answers to which can provide insights into the physics of strange metal systems of interest in condensed matter physics. What is the fate of two strange metals that are coupled to each other? Is it possible to obtain a transition from one kind of strange metal to another, and if so what is the physics of such transitions? In this paper, we address, inter alia, these questions by constructing and studying a model that involves two coupled SYK systems containing two flavors of fermions (just as in ref.~\cite{BanerjeeAltman2016}). The first, $c$-flavor  on $\Lc$ sites have $\qc$-body random interactions between them, while the second flavor of $\Psi$-fermions on separate $\LPsi$ sites have $\qPsi$ body random all-to-all interactions. We show that new physics emerges at low temperatures when these two flavors are coupled by an $r$-body all-to-all interaction between the $c$ and $\Psi$ fermions. We show several new phases emerge depending on the fraction $\fx = \LPsi/\Lc$ which include phases such as the ``\nsm '' (where one flavor is in a strange metal phase, and the other is fully gapped) , and a Mott like (both flavors fully gapped) phase that emerges due to the correlations induced by the coupling.  Not only do these results reveal the rich physics contained in SYK like models by providing a solvable example of an instability of the strange metal phase, but we also believe that our work has a much broader reach in providing a new direction to the study of strong correlation physics in a solvable model.

\paratitle{Model} Consider a system with $c$-flavor fermions that  live on $\Lc$ sites, and another  $\Psi$-flavor fermions which live on another set of $\LPsi$ sties. The $c$ fermions are individually a SYK system with $\qc$-body interactions, while the $\Psi$ fermions by themselves from a $\qPsi$-body SYK system. These two systems are coupled via an $r$-body SYK like all-to-all random interaction resulting in the Hamiltonian
\begin{widetext}
\beq\mylabel{eqn:GBA}
\begin{split}
\cH =  \sum_{i_1,\ldots,i_{\qc},j_1,\ldots,j_{q_c}} & H^c_{i_1,\ldots,i_{\qc};j_1,\ldots,j_{q_c}}c^\dagger_{i_{\qc}}\ldots  c^\dagger_{i_1 } c_{j_1} \ldots c_{j_\qc} \\
 + \sum_{\alpha_1,\ldots,\alpha_{\qPsi},\gamma_1,\ldots,\gamma_{\qPsi}} & H^\Psi_{\alpha_1,\ldots,\alpha_{\qPsi};\gamma_1,\ldots,\gamma_{\qPsi}}\Psi^\dagger_{\alpha_{\qPsi}}\ldots  \Psi^\dagger_{\alpha_1 } \Psi_{\gamma_1} \ldots \Psi_{\gamma_\qPsi} \\
 + \sum_{i_1,\ldots,i_{r},\alpha_1,\ldots,\alpha_{r}} & H^{c\Psi}_{i_1,\ldots,i_{r};\alpha_1,\ldots,\alpha_{r}}c^\dagger_{i_{r}}\ldots  c^\dagger_{i_1 }  \Psi_{\gamma_1} \ldots \Psi_{\gamma_r} + \mbox{h.~c.}
\end{split}
\eeq
\end{widetext}
where $i$s and $j$s run from $1,\ldots,\Lc$ and $\alpha$s and $\gamma$s run from $1,\ldots,\LPsi$, $c^\dagger_i$ and $\Psi^\dagger_\alpha$ are electron operators associated with the $c$ and $\Psi$ fermions. The matrix elements $H^c$, $H^{\Psi}$ and $H^{c\Psi}$ have the necessary properties demanded by fermion anti-symmetry and are Gaussian random complex variables with zero mean and respective variances $\frac{2 \Jc^2}{\qc \Lc^{2\qc-1}}$, $\frac{2 \JPsi^2}{\qPsi \LPsi^{2\qPsi-1}}$ and $\frac{V^2}{r (\sqrt{\Lc \LPsi})^{2r-1})}$. This model is specified by seven parameters -- $\qc$, $\qPsi$, $r$ and energies $\Jc$, $\JPsi$, $V$ which describe the interactions, and $\fx = \frac{\LPsi}{\Lc}$ which denotes the relative fraction of the two types of sites. The system has one conserved quantity namely the total number of particles which is captured by the $U(1)$ global phase invariance of the Hamiltonian \eqn{eqn:GBA}.

The equilibrium state of the system at a temperature $T$ and chemical potential $\mu$ (we will focus explicitly  on $\mu=0$ in this paper) is obtained by writing the partition function as a path integral over fermion fields defined over  imaginary time $\tau \in [0,\beta], \beta=1/T$. After disorder averaging the action (free energy) of the system is given by
\begin{widetext}
\beq\mylabel{eqn:Action}
\begin{split}
{\mathcal{S}} = N \Xi &= \frac{N}{1+\fx} \left[ -\frac{1}{\beta}\ln\det[-G^{-1}_c] - \frac{\fx}{\beta}\ln\det[-G^{-1}_\Psi]
  \right.- (-1)^\qc \frac{J_c^2}{2 \qc}  \int_0^\beta  \D{\tau}   G^\qc_c(-\tau) G^\qc_c(\tau) \\
  & - (-1)^{\qPsi} \fx \frac{J_\Psi^2}{2 \qPsi} \int_0^\beta   \D{\tau}   G^\qPsi_\Psi(\tau)G^\qPsi_\Psi(-\tau) -(-1)^r  \sqrt{\fx} \frac{V^2}{r} \int_0^\beta \D{\tau} 
   G_c^r(-\tau)G_\Psi^r(\tau)\\
  & \left. - \int_0^\beta \D{\tau} \, \left( \Sigc(\tau) G_c(-\tau) + \fx \SigPsi(\tau) G_\Psi(-\tau) \right) \right]
  \end{split}
\eeq
\end{widetext}
where $\Lc= \Lo$, $\LPsi= f \Lo$, and  the fermion Green's functions $G_s(\tau)$ and the self energy  $\Sigma_s(\tau)$ have been introduced. Here and henceforth $s=+1\equiv c$ and $s = -1 \equiv \Psi$.  The fermion Green's function satisfies the condition $G_s(-\tau) = - G_s(\beta - \tau)$, and the Dyson equation $G_s^{-1}(\ci \omega_n) = G_0^{-1}(\ci \omega_n) - \Sigma_s(\ci \omega_n)$ where $\ci = \sqrt{-1}$, $\omega_n = (2 n + 1) \pi T$ is a fermionic Matsubara frequency and  $G_0^{-1}(\ci \omega_n) = \ci \omega_n + \mu$. For large $N$, the saddle point values of these functions, that adequately describe the physics, satisfy the ``self consistency'' conditions
\beq\mylabel{eqn:SP}
\begin{split}
\Sigma_s(\tau)  = & (-1)^{\qs+1}  J_s^2 G_s^{\qs-1}(-\tau) G_s^\qs(\tau)  \\
& + (-1)^{r+1}  (\sqrt{\fx})^s V^2 G_s^{r-1}(-\tau) G_{\bar{s}}^r(\tau) 
\end{split}
\eeq 
where $\bar{s}=-s$. As noted above, in this paper we shall focus on the case where the chemical potential $\mu$ vanishes, i.~e., $\mu = 0$. At this value of $\mu$ the disorder averaged system has particle hole symmetry for {\em any} values of the seven parameters introduced above. The symmetry is implemented by
\beq\mylabel{eqn:PHS}
G_s(\tau) \mapsto G_s(\beta - \tau), \;\;\;\; \Sigma_s(\tau) \mapsto \Sigma_s(\beta - \tau)
\eeq
which leaves the action \eqn{eqn:Action} invariant. This particle hole symmetry is described by the group $Z_2^{\textup{ph}}$. For special values of the parameters, the system may posses further symmetries: for example, for $f=1$, when $\Jc = \JPsi$ and $\qc = \qPsi$, the system has a flavor exchange symmetry implemented via
\beq\mylabel{eqn:Flavor}
G_{s}(\tau) \mapsto G_{\bar{s}}(\tau),  \;\;\;\; \Sigma_{s}(\tau) \mapsto \Sigma_{\bar{s}}(\tau).
\eeq
This flavor symmetry is described by the group $Z_2^F$.

\paratitle{Analysis} An understanding of the states realized in model introduced above can be obtained using the ideas developed in refs.~\cite{SachdevYe1993,Kitaev2015, Sachdev2015PRX, BanerjeeAltman2016}. The central observation is that the ground state has an infrared conformal symmetry which can captured by an approximate Dyson equation as
\beq\mylabel{eqn:conformal}
\Sigma_s(\ci \omega_n) G_s(\ci \omega_n) = -1.
\eeq
The infrared physics is then captured by the ansatz
\beq
G_s(\tau) = - C_s \frac{\sgn{\tau}}{|\tau|^{2 \Delta_s}}
\eeq
where $\Delta_s$ is the dimension of the fermion field and $C_s$ a constant (both as yet undetermined). Using \eqn{eqn:SP}, we then obtain
\beq
\Sigma_s(\tau) \approx D_s \frac{\sgn{\tau}}{|\tau|^{\gamma_s}}
\eeq
where $D_s$ depends on the $C_s$s, $J_s$s and $V$, $\gamma_s = \min{(2 \Delta_s(2 q_s -1), 2 \Delta_s (r-1) + 2 \Delta_{\bar{s}} r)}$. Physically, this arises from choosing the dominant term on the right hand side of \eqn{eqn:SP}; in the case when  $2 \Delta_s(2 q_s -1) = 2 \Delta_s (r-1) + 2 \Delta_{\bar{s}} r$, then both terms in \eqn{eqn:SP} have to be treated on an equal footing. The condition \eqn{eqn:conformal} can now be used to solve for $\Delta_s$ and $C_s$. 

When $V=0$, only the first term on the rhs of \eqn{eqn:SP} is present for both flavors, and we have two uncoupled SYK systems. One finds
\beq
\Delta_s = \frac{1}{2 q_s} \equiv \Delta^0_s,\;\;\; C_s^{2 q_s} = \frac{1}{J_s^2} g(\Delta_s)
\eeq
where $g(\Delta) = \frac{1}{\pi} ( \half - \Delta) \tan(\pi \Delta)$. We now investigate ``relevance'', ``irrelevance'', or ``marginality'' of the coupling term (second term of \eqn{eqn:SP}) for each of the flavors. There are four possibilities (we take $\qc \ge \qPsi$ without loss of generality) as sketched below:

\noindent{{\bf 1.}} The coupling $V$ is marginal for both $c$ and $\Psi$ flavors. A necessary condition for this to occur is that 
\beq\mylabel{eqn:rstar}
r =  \frac{2 \qc \qPsi}{\qc + \qPsi} \equiv r_\star.
\eeq
In this case, $\Delta_s = \Delta^0_s$. We note here that when $r> r_\star$ and $V$ is ``small'' the coupling will be irrelevant for both flavors and the infrared physics is identical to the uncoupled system.

\noindent{{\bf 2.}} The coupling $V$ is marginal for $c$ fermions and relevant for the $\Psi$ fermions (which means that the second term in \eqn{eqn:SP} dominates the $\Psi$ self energy in the infrared).  Such a solution is expected for $r_\star > r > q_c$, and small values of $f$. The analysis reveals that the $\Delta_c = \Delta^0_c$ while, $\Delta_\Psi = \frac{1}{r} - \Delta^0_c$, and $C_c^{2 \qc} = \frac{1}{\Jc^2}(g(\Delta^0_c) - \fx g(\Delta_\Psi)$ (a similar expression can be found for $C_\Psi$). An important point to note that this is indeed an infrared solution, in that the conditions are met only when the frequency $\omega \lesssim \omega_{\textup{cut}}$ which depends on all the parameters (including $V$ and $f$; we do not give the expression for $\omega_{\textup{cut}}$ here). The fact that $\omega_{\textup{cut}}> 0$ provides a condition on $f$ for the existence of this solution, $f < \frac{g(\Delta^0_c)}{g(\Delta_\Psi)} = f_l$.

\noindent{{\bf 3.}} Coupling is marginal for the $\Psi$ flavor while relevant for the $c$ flavor. Such a situation is expected for $f \gg 1$. The analysis proceeds in exactly the same fashion as the previous case, and one obtains similar expressions for fermion dimensions and $C$s with the roles of $c$ and $\Psi$ reversed. Again, such an infrared solution is valid only when $f >\frac{g(\Delta_c)}{g(\Delta^0_\Psi)} = f_h$.

%\onecolumngrid
\begin{figure*}
\includegraphics[width=\textwidth]{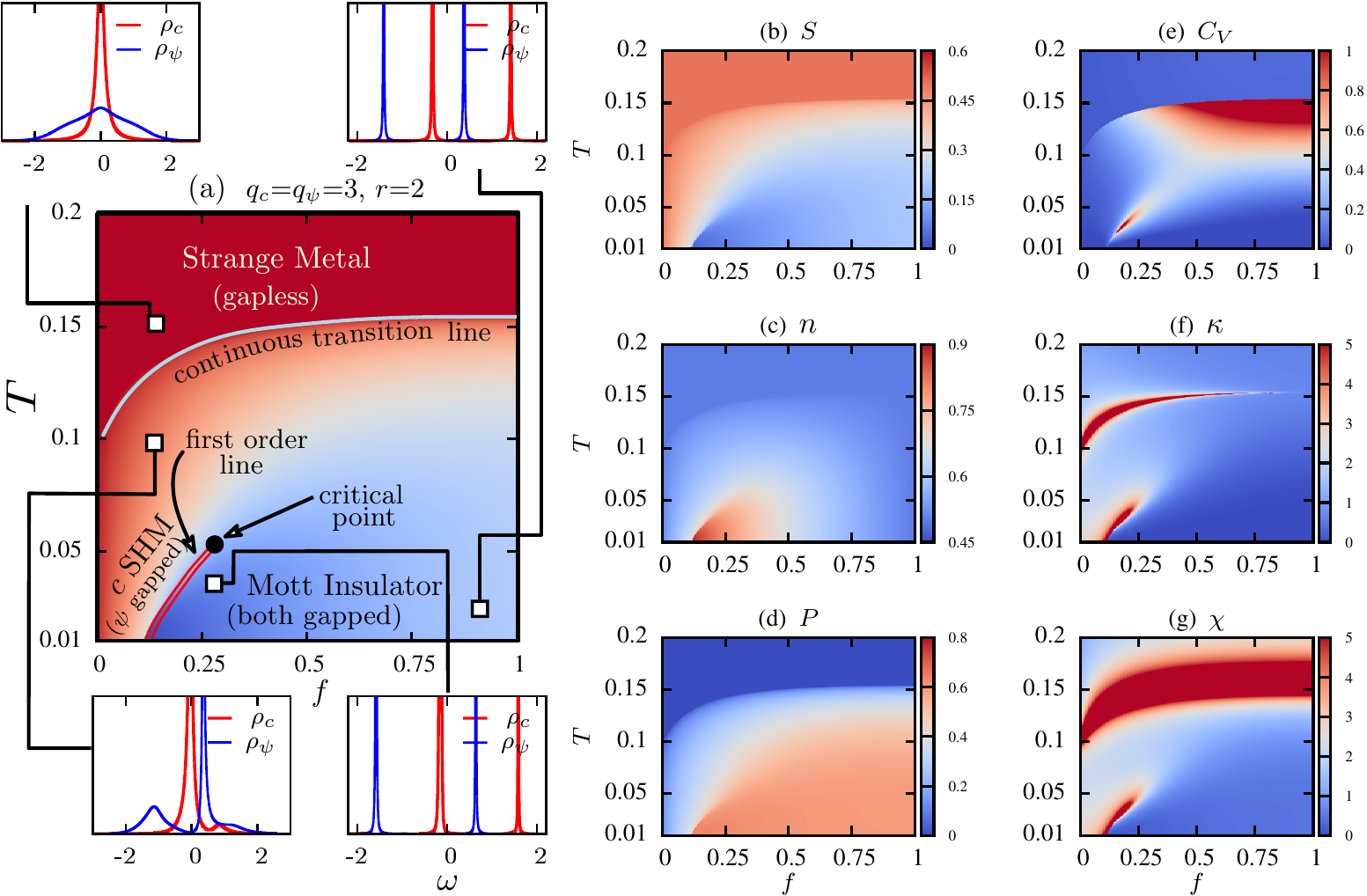}
\caption{{\bf Numerical results for $\qc=\qPsi=3,r=2$, $\Jc=\JPsi=V=1$.}
{\bf(a)} Phase diagram and spectral functions (SHM stands for  strange half metal),  {\bf (b) -- (f):} Plots of entropy $S$, density $n$, polarization $M$, specific heat $C_V$, compressibility $\kappa$, and susceptibility $\chi$ in the $T-f$ plane.}
\mylabel{fig:R332}
\end{figure*}
%\twocolumngrid

\noindent{{\bf 4.}} The coupling is relevant for both flavors. This is expected when $r < \qPsi$, and for ``large'' $V$. In this case, it can be shown that $\Delta_c + \Delta_\Psi = \frac{1}{r}$, and $\Delta_c $ is an $f$ dependent quantity that is determined by the solution of $f = \frac{g(\Delta_c)}{g(\frac{1}{r} - \Delta_c)}$. The fermion dimensions are $f$ dependent, i.e., one can continuously change the fermion dimension by changing $f$! In other words, the character of the emergent strange metals is determined by the parameter $f$. 

To conclude the discussion, one can obtain conformal solutions to the ground state in different conditions, and when this is done the solution will be either the uncoupled SYK state of the two flavors, or one of the four other states listed above. An important point to be noted is that there are regimes of parameters (for example $f_l < f < f_h$) for which we have not been able to find a conformal solution. The physical implication of this is that for $\qPsi < r < r_\star$, the system is a strange metal with dominant $c$ fermions at small $f$, and transits to another strange metal dominated by $\Psi$ fermions with increasing $f$. In other words, one can tune a transition from one type of strange metal to another by tuning $f$. Note that these two strange metal phases are separated by a ``nonconformal'' phase (for $f_l < f < f_h$). Finally, the strange metal phases captured by the conformal solution will have a finite ground state entropy, and a concomitant high entropy state at nonzero temperatures.

\paratitle{Numerics} Primarily motivated by the urge to understand the nature of the ``nonconformal'' states that lie in between the strange metal phases at small and large $f$, we performed calculations that solve the self energy equations \eqn{eqn:SP} numerically. To characterize the states obtained we have calculated several quantities, including entropy $S$, the specific heat $C_V$, the density of particles $\dens$, the ``polarization'' $P = \dens_c - \dens_\Psi$($n_s=G_s(\tau=0^-)$), the compressibility $ \left. \kappa = \dou \dens /\dou \mu \right|_{\mu=0}$, and the ``magnetic susceptibility'' $\chi$ (calculated by introducing a magnetic field ($h$) coupling to the two fermion flavors with opposite signs and then taking  $\partial^2 \Xi/\partial h^2|_{h=0}$). We have also obtained the spectral functions $\rho_c(\omega)$ and $\rho_\Psi(\omega)$ of the $c$ and $\Psi$ fermions where $\omega$ is the frequency.

We begin the discussion with the case with $\qc=\qPsi=3, r=2$, and $\Jc=\JPsi=V=1$. One expects a conformal solution of type 4 discussed above. \Fig{fig:R332} shows the results of the numerical calculations for this case. For temperatures $T \gtrsim 0.15$, a high entropy state consistent with the conformal solution ({\bf 4}) is indeed found (detailed comparison with the analytics will be presented elsewhere) for all values of $f$. The density $\dens$ of particles per site in this regime is $0.5$ as this state is particle-hole symmetric.  The spectral functions of both $c$ and $\Psi$ fermions show the metallic character of this state.

 \begin{figure*}
 \includegraphics[width=\textwidth]{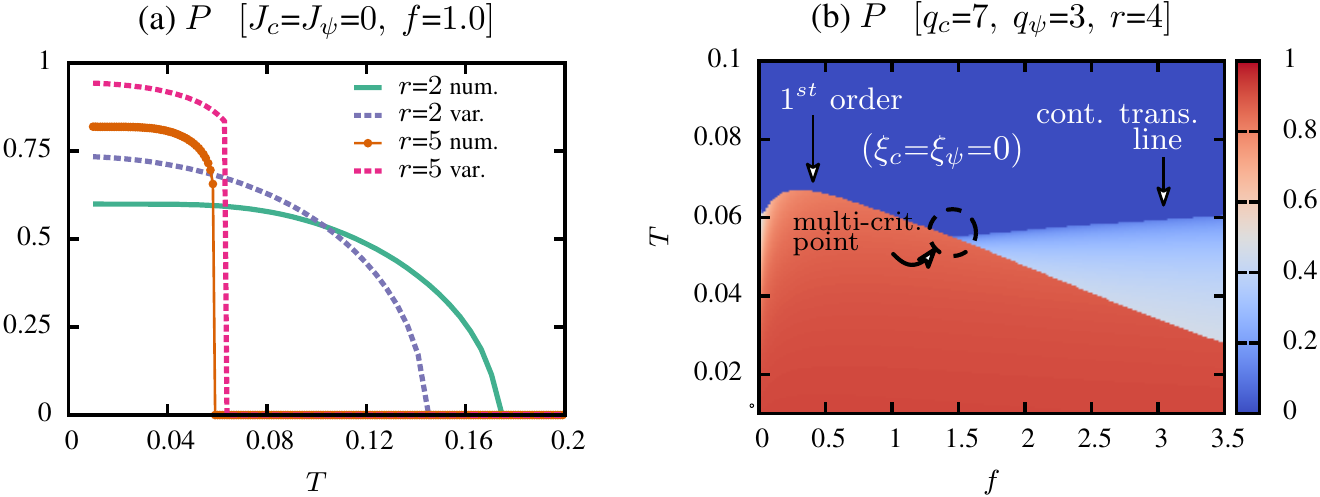}
 \caption{{\bf Comparison of variational results with full numerics:} 
 {\bf (a)} Plot of polarization($P$=$n_c-n_\psi$) obtained via numerical (solid lines) and variational(dashed lines) approaches for $r$=$2,5$ and $Jc$=$J_\psi$=$0$.
  {\bf (b)} Phase diagram of polarization in the $T-f$ plane obtained using the variational approach for $q_c$=$7$, $q_\psi$=$3$, $r$=$4$ and $J_c$=$J_\psi$=$V$=$1$.}
 \mylabel{fig:var}
 \end{figure*}

Things, however, take a dramatic turn at lower temperatures. For all $f$ in the range $0 < f < 1$ (physics of $f>1$ can be obtained by replacing $f \mapsto 1/f$), we find that the system encounters a continuous phase transition at a temperature $T_c(f)$ which depends on $f$ (see \Fig{fig:R332}({\bf a})).  The state for $T < T_c(f)$ at arbitrary $f$ breaks the particle-hole symmetry, as is immediately evident from the fact that $n$ deviates from $0.5$. Also, the polarization $P$ is nonzero in the regime $T < T_c(f)$. The transition is continuous for $0 < f < 1$ and is Landau like as we have confirmed by a study of free energy which is continuous, the specific heat $C_V$ which shows a jump, susceptibility $\chi$ which behaves as $1/|T - T_c|$ etc. The entropy $S$ begins to fall  with decreasing $T$ (for $T< T_c$). Most interestingly, the states realized below $T_c$ are very different at different $f$s. For $f \ll 1$, we obtain a state with a high entropy density, and finite compressibility. A study of the spectral function (see \Fig{fig:R332}({\bf a})) shows that in this regime, the $c$ fermions are gapless with a diverging density of states at $\omega =0$, while the $\Psi$ fermions are gapped! We dub this state a \nsm\ (\NSMab), and to make it explicit that the $c$-Fermions are gapless, we denote this by $c$-\NSMab\ state. The choice of this name is motivated by physics seen in certain correlated materials\cite{Groot1983PRL}.
On the other hand, the low temperature state at $f \approx 1$ has both fermions fully gapped. The spectral functions of both fermions show two peaks with a vanishing (exponetially small) spectral weight at the chemical potential.
This state has vanishing entropy and a vanishing compressibility. All of these indicate a state akin to the Mott insulator (MI).
 At low temperatures, upon increasing $f$ starting from $f \approx 0$, the $c$-\NSMab\ state encounters a first order transition and transits to a MI like state\footnote{The usual Mott insulator, encountered for example in the Hubbard model, occurs at half filling. While the state at $f=1$ is indeed at half filling $n=0.5$, the density of particles for $f < 1$ does not correspond to half filling. We, nevertheless, adopt the name Mott insulator as the interaction gaps out both the fermions.}.

  Intriguingly, this first order transition ends at a critical point at a certain value of temperature and $f$ as indicated in the \Fig{fig:R332} ({\bf a}) and many thermodynamic quantities diverge near this critical point. In hindsight, given the existence of the first order transition between the $c$-\NSMab\ state and the MI state at low temperature, a critical point is expected to occur at higher temperatures as the two states $c$-\NSMab\ and MI both have the same symmetries. Note also that for any $T$ and $f$ there is another distinct state (than the one shown in \fig{fig:R332}) with same free energy that is particle hole transformation related -- this state will have a density of $1-n$ and a polarization of $-P$ corresponding to the one shown in the figure. Finally, we note that at $f=1$ the system has a $Z_2^{\textup{ph}} \times Z_2^F$ symmetry. Below $T_c$, this symmetry is broken down to another smaller $Z_2$ symmetry described by $G_{s}(\tau) \mapsto G_{\bar{s}}(\beta -\tau), \Sigma_{s}(\tau) \mapsto \Sigma_{\bar{s}}(\beta -\tau)$. Consequently $n$ remains $0.5$ at all temperatures, and the state only develops a polarization $P$ for $T < T_c$.

\paratitle{Discussion} We now address the physics that leads to the results obtained in the last section. A good starting point for this purpose is to set $\Jc = \JPsi =0 $, we thus have a system where the sole interaction is the $r$-body term between the $c$ and $\Psi$ fermions. The physics of this system (at the saddle point level) can be viewed a theory of two classical fields $G_s(\tau)$ ($s=c,\Psi$) defined on an interval of length $\beta$. These two fields have a ``long range interaction'' as embodied in the term  $-  \sqrt{\fx} \frac{V^2}{r} \int_0^\beta \D{\tau} G_c^r(\beta-\tau)G_\Psi^r(\tau)$ (see \eqn{eqn:Action}). The interaction is ``long ranged'' as the $G_c(\tau)$ couples to $G_\Psi(\beta - \tau)$ with a strength of order $V^2$. In a particle hole symmetric metallic state $G_s(\beta -\tau) = G_s(\tau)$, the spectral function is symmetric $\rho_s(-\omega) = \rho_s(\omega)$ and thus contributes to the reduction in free energy via its higher entropy. On the other hand, such a state is not best one to optimize the long interaction term just discussed; a state where the $Gs$ are not symmetric about $\tau = \beta/2$ would be preferred to gain this long range interaction energy. Since the interaction term begins to dominate at lower temperatures, one might expect the breaking of particle hole symmetry at lower temperatures by going to states $G_s(\beta - \tau) \ne G_s(\tau)$. In other words, the system favors to have a majority of one flavor of particles -- much like what happens in system with Mott physics.

%\onecolumngrid
\begin{figure*}
\includegraphics[width=\textwidth]{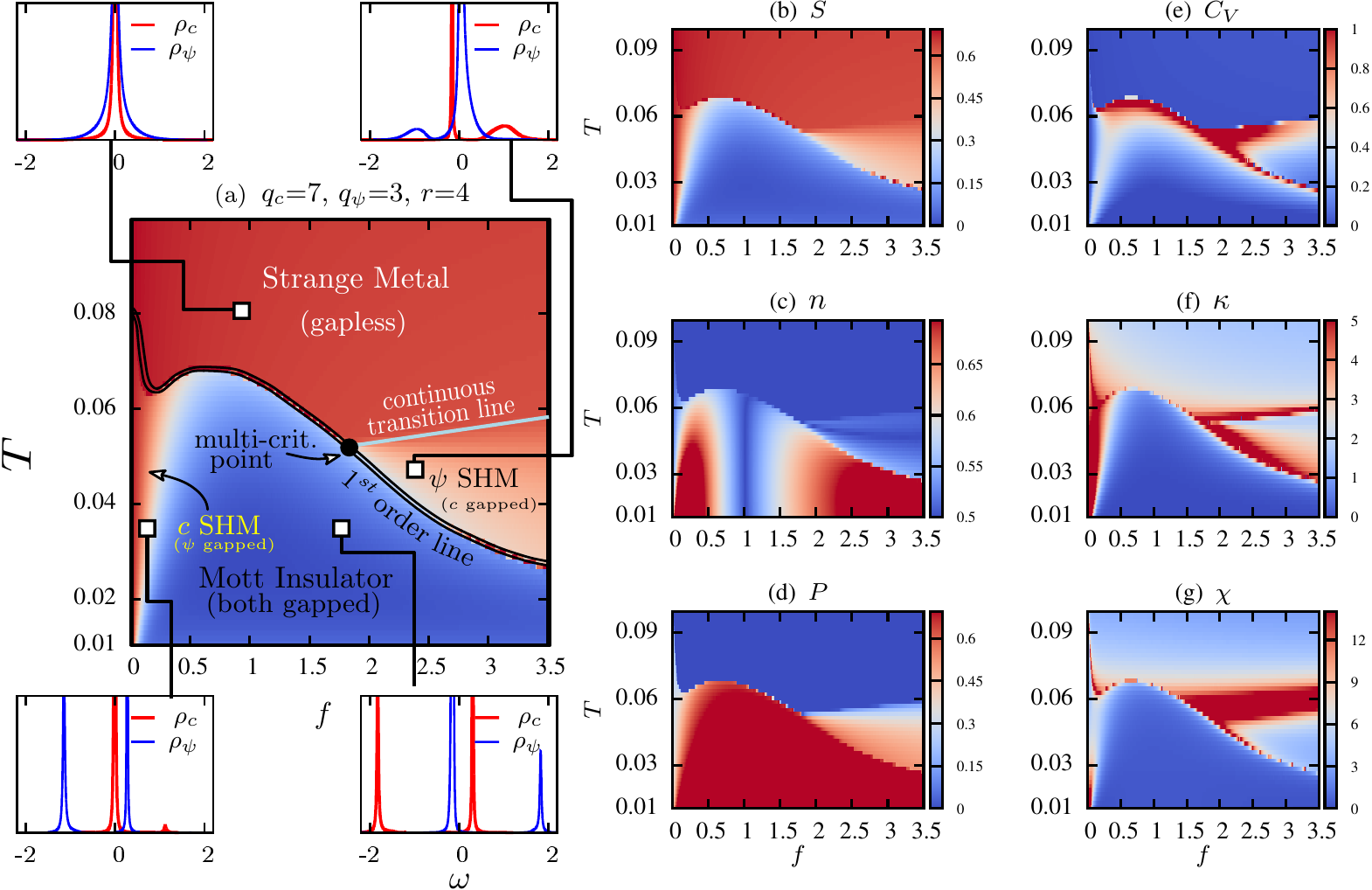}
\caption{{\bf Numerical results for $\qc=7,\qPsi=3,r=4$, $\Jc=\JPsi=V=1$.} {\bf(a)} Phase diagram and spectral functions,  {\bf (b) -- (f):} Plots of entropy $S$, density $n$, magnetization $M$, specific heat $C_V$, compressibility $\kappa$, and susceptibility $\chi$ in the $T-f$ plane.}
\mylabel{fig:R734}
\end{figure*}

These qualitative ideas can be cast on firmer footing, by using them in developing a variational ansatz to find the equilibrium state. Focusing on $f=1$, we introduce particle hole symmetry breaking by the ansatz $G_s(\ci \omega_n) = \frac{1}{\ci \omega_n + s \xi}, \Sigma_s(\ci \omega_n) = - s \xi$ where $\xi$ is a variational energy scale. The variational free energy then takes the form
\begin{widetext}
\beq
\Xi_{\textup{var}} = \frac{1}{2} \left(-\frac{V^2 \sinh \left(\frac{\xi  r}{T}\right) \left(\frac{1}{2 \cosh \left(\frac{\xi
   }{T}\right)+2}\right)^r}{\xi  r^2}-T \left(\log \left(e^{-\frac{\xi }{T}}+1\right)+\log \left(e^{\frac{\xi
   }{T}}+1\right)\right)+\xi  \tanh \left(\frac{\xi }{2 T}\right)\right).
\eeq
\end{widetext}
Analyzing this variational free energy one arrives at several conclusions: (i) For $r=1$, $\xi =0$ is always the minimum of the free energy and a particle-hole symmetric state is always favored. (ii) Transitions are possible (a nonzero $\xi$ minimizes $\Xi_{\textup{var}}$) when $r \ge 2$ at sufficiently low temperatures. (iii) The nature of the transitions depend on the value of $r$. In fact, we find that for $r = 2, 3$ the transition is continuous, while for $r \ge 4$ the transition is first order!

Quite encouragingly, these predictions of the simple variational approach are indeed borne out by the full numerical solution; one does have a continuous transition for $r=2$ while it is first order for $r\geq 4$ (see \fig{fig:var}(a)). Of course, the $T_c$ is only approximately reproduced by the variational approach. These positive results have motivated us to write a variational free energy for any arbitrary set of seven parameters of the system which depends on two variational parameters $\xi_c$ and $\xi_\Psi$. We used this to study the phase digram of the system in the $T-f$ plane for $\qc=7$, $\qPsi=3$, $r=4$ with $\Jc=\JPsi=V=1$. A remarkably rich phase diagram(see \fig{fig:var}(b)) with multi-critical points etc.~is predicted within the variational framework.

%\twocolumngrid 

\paratitle{Further Results} Armed with the insights obtained from the variational approach, we performed detailed numerical calculations for the case with $\qc=7,\ \qPsi=3,\ r=4$ and $\Jc=\JPsi=V=1$ whose results are shown in \Fig{fig:R734}. One again finds a phase with a large entropy density at high enough temperature that corresponds to the conformal solutions discussed previously. Phase transitions are encountered at lower temperatures. The particle hole symmetry breaking transition is first order for $f$ less than that corresponding to the multi-critical point at $f \approx 1.75$. For larger values of $f$, the particle hole symmetry is broken via a continuous transition. Below $T_c$, at small $f$ there is a $c$-\NSMab\ phase, which in this case is smoothly connected to the Mott insulating phase(see \fig{fig:R734}({\bf a})) to the lowest temperatures that we have access to. For $f \gtrsim 1.75$, there is $\psi$-\NSMab\ phase which is separated from the MI phase via a first order line (which ends at the multi-critical point). Quite interestingly, the variational approach produces a qualitatively similar picture for this case(see \fig{fig:var}(b)). We note here that while the overall structure of phase diagram for the case discussed in \fig{fig:R332} is also reproduced by the variational approach, the are differences at smaller $f$.

\paratitle{Summary and Perspective} In this paper we have proposed and studied a model with two flavors of fermions each of which form an SYK system and interact with each other by an all-to-all $r$ body  interaction. In addition to the strength of this $r$-body interaction, the ratio $f$ of the number of sites for each flavor is a crucial parameter. A key conclusion is that when $r \ge 2$, the coupling can produce drastic changes in the ground state producing new phases like the \NSM\ (\NSMab)\ and Mott Insulating (MI) phases. The generic scenario is that  strange metal phases (described by a conformal solution) found at high temperatures undergo a phase transition (which can be either continuous or first order) to a phase where one or both flavors are gapped by the breaking of particle hole symmetry. At a fixed low temperature, one obtains a \NSMab\ state for small or large $f$, and this state is typically separated from a Mott insulating phase via a first order transition.

These findings provide several directions for further investigation. It will be interesting to explore the connection of this novel physics to the gravitational view of the SYK model. From a condensed matter perspective, this could also provide a new framework for the study  of strong correlations and Mott physics by generalizations of this model to higher dimensions, and more generically, the instabilities of the strange metal phase.

\bigskip 
\paratitle{Acknowledgements} The authors thank Subir Sachdev for discussions,  and Sumilan Banerjee for discussions and comments on the manuscript. Financial support from CSIR (AH), DST(VBS) is acknowledged.

%%%%%%%%%%%%%%%%%%%%%%%%%%%%%%%%%%%%%%%%%%%%%%%%%
%% BIB FILE
%%\bibliographystyle{naturemag.bst}
%%\bibliography{refAllnature}   %Nature journal doesn't have URLS for the articles. %%So %this refAllnature 
%%clearpage                    %file has all URLs removed from bibtex entries using %%sed.
%%%%%%%%%%%%%%%%%%%%%%%%%%%%%%%%%%%%%%%%%%%%%%%%

\bibliography{refAll} % UNCOMMENT FOR PRL FORMAT
%% \iffalse              % UNCOMMENT FOR PRL FORMAT

%\def\makeSM{1}
\ifdefined\makeSM

%%%%%%%%%%%%%%%%%%%%%%%%%%%%%%%%%%%%%%%%%%%5
\newwrite\tempfile
\immediate\openout\tempfile=junkSM.\jobname
\immediate\write\tempfile{\noexpand{\thepage} }
\immediate\closeout\tempfile

\clearpage

\newpage

%%%%%%%%%%%%%%%%%%%%%%%%%%%%%%%%%%%%%%%%%%%%%%%%%%%%%%%%%%%%%%%%%%%%%%%%%%%%%%%%%%%%%%%%%%%%%%%%%%%

%\end{document}
%UNCOMMENT THIS TO GENERATE FILE WITHOUT SUPPELEMENTAL MATERIAL

\appendix

\renewcommand{\appendixname}{}
\renewcommand{\thesection}{{S\arabic{section}}}
\renewcommand{\theequation}{\thesection.\arabic{equation}}
 
\setcounter{page}{1}
\setcounter{figure}{0}

\widetext

\centerline{\bf Supplemental Material}
\centerline{\bf for}
\centerline{\bf \titlename}
\centerline{by Arijit Haldar and Vijay B.~Shenoy}
\author{Arijit Haldar}
\email{arijit@physics.iisc.ernet.in}
\author{Vijay B. Shenoy}
\email{shenoy@physics.iisc.ernet.in}
\affiliation{Centre for Condensed Matter Theory, Department of Physics, Indian Institute of Science, Bangalore 560 012, India}
%\bigskip
%\end{widetext}
%\medskip

%%%%%%%%%%%%%%%%%%%
%\setcounter{equation}{0}
%\setcounter{figure}{0}
%\setcounter{table}{0}
%\setcounter{page}{1}
%\makeatletter
%\renewcommand{\theequation}{S1.\arabic{equation}}
%\renewcommand{\thefigure}{S1.\arabic{figure}}

\section{Large $N$ formulation}

\section{Conformal States}

\section{Variational Analysis}

\section{Calculation of the Spectral Function}

%\section{}

\clearpage

\fi %%TO MAKE SUPPLEMENTAL MATERIAL

\end{document}